\documentclass[sigconf]{acmart}
\usepackage{multirow}
\usepackage{algpseudocode}
\usepackage{algorithm}
\usepackage{mwe}
\usepackage{graphicx}
\usepackage{comment}
\usepackage{booktabs} % 导入booktabs宏包用于更好的表格线条
\usepackage{colortbl}
\usepackage{arydshln}
\usepackage{subfig}
\usepackage{caption}
\usepackage{enumitem}
\usepackage{placeins}
\AtBeginDocument{%
  \providecommand\BibTeX{{%
    \normalfont B\kern-0.5em{\scshape i\kern-0.25em b}\kern-0.8em\TeX}}}

\setcopyright{acmcopyright}
\copyrightyear{2018}
\acmYear{2018}
\acmDOI{XXXXXXX.XXXXXXX}

\acmConference[xx]{Make sure to enter the correct
  conference title from your rights confirmation emai}{July 14--18,
  2024}{Washington D.C.}
\acmBooktitle{Woodstock '18: ACM Symposium on Neural Gaze Detection,
 June 03--05, 2018, Woodstock, NY} 
\acmPrice{15.00}
\acmISBN{978-1-4503-XXXX-X/18/06}

\begin{document}

\title{Reinforcement Learning-based Recommender Systems with Large Language Models for State Reward and Action Modeling
}
%\titlerunning{Reinforcement Learning-based Recommender Systems with LLMs for State Reward and Action Modeling}

%\title{Grounding Large Language Models as User Feedback %Augmented Environment %Efficiently 
%in Reinforcement Learning based Recommender Systems%: On Environment and Agent
%Guiding Reinforcement Learning based Sequential Recommendation Efficiently with Large Language Models: On Environment and Agent
%%}

\author{Jie Wang}
\affiliation{%
  \institution{University of Glasgow}
  \country{}
}
\email{j.wang.9@research.gla.ac.uk}

\author{Alexandros Karatzoglou}
\affiliation{%
  \institution{Google Research, London}
  \country{}
}
\email{alexandros.karatzoglou@gmail.com}

\author{Ioannis Arapakis}
\affiliation{%
  \institution{Telefonica Research, Barcelona}
  \country{}
}
\email{ioannis.arapakis@telefonica.com}

\author{Joemon M. Jose}
\affiliation{%
 \institution{University of Glasgow}
 \country{}
}
\email{joemon.jose@glasgow.ac.uk}

\begin{abstract}

Reinforcement Learning (RL)-based recommender systems have demonstrated promising performance in meeting user expectations by learning to make accurate next-item recommendations from historical user-item interactions. However, existing offline RL-based sequential recommendation methods face the challenge of obtaining effective user feedback from the environment. Effectively modeling the user state and shaping an appropriate reward for recommendation remains a challenge. In this paper, we leverage language understanding capabilities and adapt large language models (LLMs) as an environment (LE) to enhance RL-based recommenders. The LE is learned from a subset of user-item interaction data, thus reducing the need for large training data, and can synthesise user feedback for offline data by: (i) acting as a state model that produces high quality states that enrich the user representation, and (ii) functioning as a reward model to accurately capture nuanced user preferences on actions. Moreover, the LE allows to generate positive actions that augment the limited offline training data. We propose a LE Augmentation (LEA) method to further improve recommendation performance by optimising jointly the supervised component and the RL policy, using the augmented actions and historical user signals. We use LEA, the state and reward models in conjunction with state-of-the-art RL recommenders and report experimental results on two publicly available datasets.

\end{abstract}


%\ccsdesc[500]{Computer systems organization~Embedded systems}
%\ccsdesc[500]{Diversity and Novelty in recommendation}


 % \caption{Seattle Mariners at Spring Training, 2010.}
 % \Description{Enjoying the baseball game from the third-base

% \received{20 January 2023}

\maketitle

\section{Introduction}

Recommender Systems (RS) have become an essential tool 
that navigates through extensive data to deliver relevant and engaging content to users \citep{hu2018reinforcement, yuan2020} in commercial platforms. 
Sequential or next-item recommendation~\cite{hidasi2015session,tang2018personalized,yuan2019simple,kang2018self} have gained prominence, especially in music and video streaming RS,  
to recommend the next relevant item based on user-item interactions within a recent active session. Typically, sequential recommendation models~\cite{hidasi2015session}, such as those based on gated recurrent units (GRU), convolutional neural networks (CNN)~\citep{yuan2020}, and Transformer~\citep{kang2018self}, have been trained to predict the next interacted items based on historical user data in a self-supervised manner. 
To address sub-optimal recommendations due to the dependence on supervised learning, self-supervised reinforcement learning (SSRL) has been proposed. This approach trains an RL agent to satisfy user expectations, e.g., the desire for diverse content, by employing sequential models with rewards tailored to accommodate various behaviours.~\cite{xin2020self,stamenkovic2022choosing}. However, recent efforts~\cite{xin2020self,xin2022supervised,ren2023contrastive} to develop off-policy/offline RL policies trained on historical user data have been met with the challenge of constructing a high-quality environment that provides meaningful user feedback, e.g., state representation and reward function.

Large Language Models (LLMs) with knowledge-transferring capabilities have recently
received significant attention in RS~\cite{he2023large,zhang2023user,du2023enhancing}. 
In the context of sequential recommendation, LLMs have been shown to be acceptable zero-shot~\cite{bao2023tallrec} or pre-trained~\cite{geng2022recommendation} RS. These LLM-recommenders have been proven to perform on par with, or even outperform, conventional models.  
However, many works have heavily focused on fine-tuning/pre-training with user data to adapt the LLM to new recommenders, resulting in larger models with orders of magnitude more parameters
than traditional ones. Consequently,  
this necessitates committing substantial computational resources for training and
posing a challenge in effectively adapting LLMs to the sequential recommendation task.

Motivated by their generative and language understanding capabilities, we propose adapting LLM as an environment (LE) to model user behaviour and return feedback for training the RL-based recommenders. 
More importantly,
LE can help train leaner and more adaptable sequential recommendation models that outperform those trained with standard techniques without incurring additional computational costs at inference time.  
Specifically, we address the following limitations of the above-mentioned state and reward problems for RL-based RS:
RL-based recommendation approaches typically use state representations coming from a generative sequential models such that is the output/hidden state of a generative sequential model (e.g. transformer, RNN) feed with the user-logged actions that are also used to generate the recommendations. 
LLMs have been shown to be world models~\cite{Gurnee2024language,li2023emergent}, which may help generate more accurate representations of user state given the sequence of historical user actions. Additionally, in typical RL-based recommenders rewards are usually predefined w.r.t behaviour categories ~\cite{xin2020self}, e.g., purchase and click. However, a simple uniform reward setting might not accurately reflect user satisfaction across items, resulting in an agent that is unable to capture latent differences between actions.
LLMs have been shown to be capable of generating good reward estimates~\cite{kwon2023reward} that can be used to train RL algorithms.
To this end, we capitalize on these powerful LLM properties to learn an environment (LE) that acts as state and reward model to return high-utility feedback for training RL-based recommendation models. 

To construct the LE we fine-tune the LLM by introducing an item tokenization strategy, using autoregressive training. We first learn semantically-rich tokens by using the items' textual descriptions. We subsequently fine-tune the LLM  through a small subset of user data and adapters to obtain the reward model (RM) and state model (SM). Specifically, we prompt the LLM with instructions based on user-item token interactions to output the scalar reward by a score head. We then learn effective state representations by contrasting the user-item tokens interactions with the positive and negative actions. In our modular architecture, RM and SM constitute the LE that enables the acquisition of the state representations and scalar rewards for the RL-based recommendation model.

Moreover, in an offline setting, the agent is trained on fixed historical user-item interactions without probing the environment. Thus, we further propose an LE Augmentation (LEA) method prompting the obtained LE to enrich the offline data for RL-based sequential recommendation. In particular, LE is tasked with selecting potentially positive feedback by prompting it with a combination of user historical behavior and a sampled list of items. 
This step aims at leveraging the predicted positive items as positive samples to augment the training of the supervised learning component, and as 
positive actions to reinforce the training of the RL agent.

Finally, we  train the supervised loss and the RL loss over the original historical user data and the augmented positive samples. At inference stage, only the sequential model with the supervised head is used for evaluation of the top-$k$ recommendation performance, to guarantee efficiency. To validate the effectiveness of our method,
we compare LEA with two state-of-the-art Q-value-based RL frameworks, with two sequential models as the backbone. 
We also apply directly LE to the above frameworks by enhancing the state representation and reward function, and demonstrate  
significant performance gains on two publicly available datasets. 

Our contributions can be summarised as follows:
 \begin{itemize}[leftmargin=*]
    \item We propose an LLM-based Environment (LE), acting as the state model and reward function, to improve the performance of the offline RL-based recommender systems. 

    \item We present an efficient fine-tuning method for adapting LLMs for LE using limited user data. Additionally, we propose an item-tokenization strategy to incorporate user data and improve training efficiency.
    
    \item We introduce a positive feedback augmentation approach LEA to enhance both supervised learning and Q-learning. The LE is utilized as the behaviour policy to infer positive signals from historical user data. 
    
    \item  We apply the environment LE and augmentation method LEA to two state-of-the-art RL-based sequential recommendation models. Experimental results on real-world datasets show a general improvement across recommendation performance metrics. Code will be publicly available upon acceptance.
 \end{itemize}

\section{Related Work}

\subsection{Supervised Reinforcement Learning for Sequential Recommendation} 

Deep learning-based sequential recommender systems (DSRS) model users' historical interactions as next-item recommendation tasks to predict their future preferences. One of the first models was proposed by~\citet{hidasi2015session} utilizing Gated Recurrent Units (GRU) to model user sequences. Subsequently, 
Convolutional neural networks (CNN)~\citep{yuan2019simple} and the Transformer architecture~\citep{tang2018personalized} were also adopted for the task. In an offline setting, the sequential recommendation problem can be viewed as a Markov Decision Process (MDP)~\cite{zhao2018recommendations,moling2012optimal,wang2020kerl}, a framework often employed in e-commerce scenarios with reinforcement learning (RL).
~\citet{hong2020nonintrusive}
integrated wireless sensing to RL Monte Carlo tree search algorithm to improve music recommendation performance. ~\citet{wang2020kerl} introduced a Knowledge-guided RL framework that enriches the environment by a Knowledge Graph.
Moreover, ~\citet{xin2020self} introduced a self-supervised reinforcement learning framework for sequential recommendation (SRLSR). The Sequential Q-Network (SQN)~\citep{xin2020self} and Soft Actor-Critic (SAC)~\citep{xin2022supervised} architectures combine a Double Q-learning head~\citep{hasselt2010double} with supervised learning to enhance the baselines accuracy with respect to user clicks and purchases. They further enhanced the frameworks with SNQN and SA2C~\cite{xin2022supervised} by using a sampling strategy to integrate negative feedback. More recent research~\cite{xin2022rethinking} proposed a paradigm of modeling the desirable cumulative rewards rather than the expected returns at each timestamp to guide the training. In other recent work, ~\citet{ren2023contrastive}  augmented the original 
states to improve SQN by contrastive learning.  

\subsection{LLMs for Sequential Recommendation}

LLMs 
~\cite{zhao2023survey,liu2023pre,min2023recent} pre-trained on massive natural language datasets with continuously enhanced transfer capabilities have increasingly garnered attention in the field of RS~\cite{chen2023large,lin2023can,fan2023recommender}. Existing adaptations of LLMs for recommendation tasks involve primarily training the LLM to be a new recommender through pre-training~\cite{geng2022recommendation}, fine-tuning~\cite{min2023recent}, prompt-based tuning~\cite{liu2023pre,ouyang2022training} etc.
For sequential recommendation, ~\citet{geng2022recommendation} pre-train T5~\cite{raffel2020exploring} with four tasks,  e.g., rating, explanation, review, and direct recommendation, for new domains in RS.
Other recent works~\citep{hou2022towards,wang2022transrec} trained recommenders through user-item interactions based on the item features extracted from BERT~\cite{devlin2018bert} in the source domain and showed encouraging performance for cross-domain recommendation. ~\citet{rajput2023recommender} generates the semantic IDs for item representations by Sentence-T5~\cite{ni2021sentence} and then predicts the semantic IDs of the next item autoregressively.  ~\citet{bao2023tallrec} efficiently fine-tunes LLaMA-7B~\cite{touvron2023llama} with LoRA adapters~\cite{hu2021lora} by an instruction prompt including item text descriptions to realize few-shot recommendations. However, these methods of employing LLMs as recommenders come with significant pre-training costs or face difficulties in preserving user signals within lengthy text sequences. Moreover, research efforts that aim to harness LLMs efficiently to improve the performance of existing ID-based sequential models have been sparse. 
Recent work~\cite{kwon2023reward} has shown that using LLM as reward model outperforms learned rewards in the context of RL applications in games.

This work diverges from existing approaches that focus solely on training new LLM-based sequential recommenders with heavy computation at training and particularly at inference time. Instead, we aim at efficiently adapting LLMs to serve as an environment component within an RL framework, with the objective of augmenting the performance of existing recommenders. 

\section{Method}%
\label{method}
\subsection{Task Formulation}
\label{task_formulation} 
\begin{figure*}[!t]
    \captionsetup[subfloat]%{captionskip=-5pt,nearskip=0pt,farskip=0pt}
    {}
    \centering
    \begin{minipage}[c]{0.47\textwidth}
    \centering
        \subfloat[ Traditional SSRL4R ]{%
        \includegraphics[trim={0 0 0 0}, clip, scale=0.2]{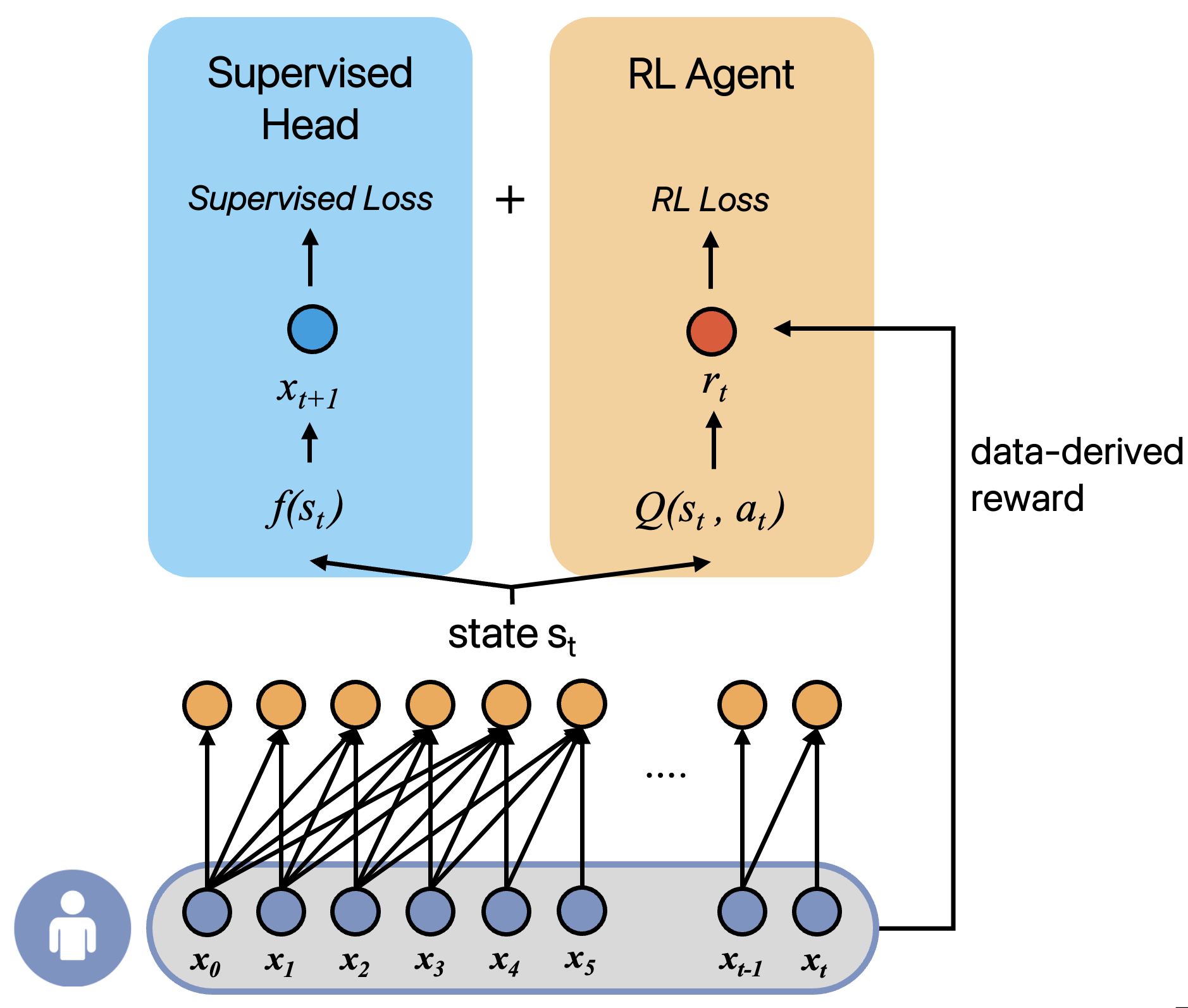}
        \label{figure1_a}
        }
    \end{minipage}
    \hspace{0.5mm}
    \begin{minipage}[c]{0.47\textwidth}
    \centering
        \subfloat[SSRL4R with state \& reward models]{%
        \includegraphics[trim={0 0 0 0}, clip, scale=0.2]
        {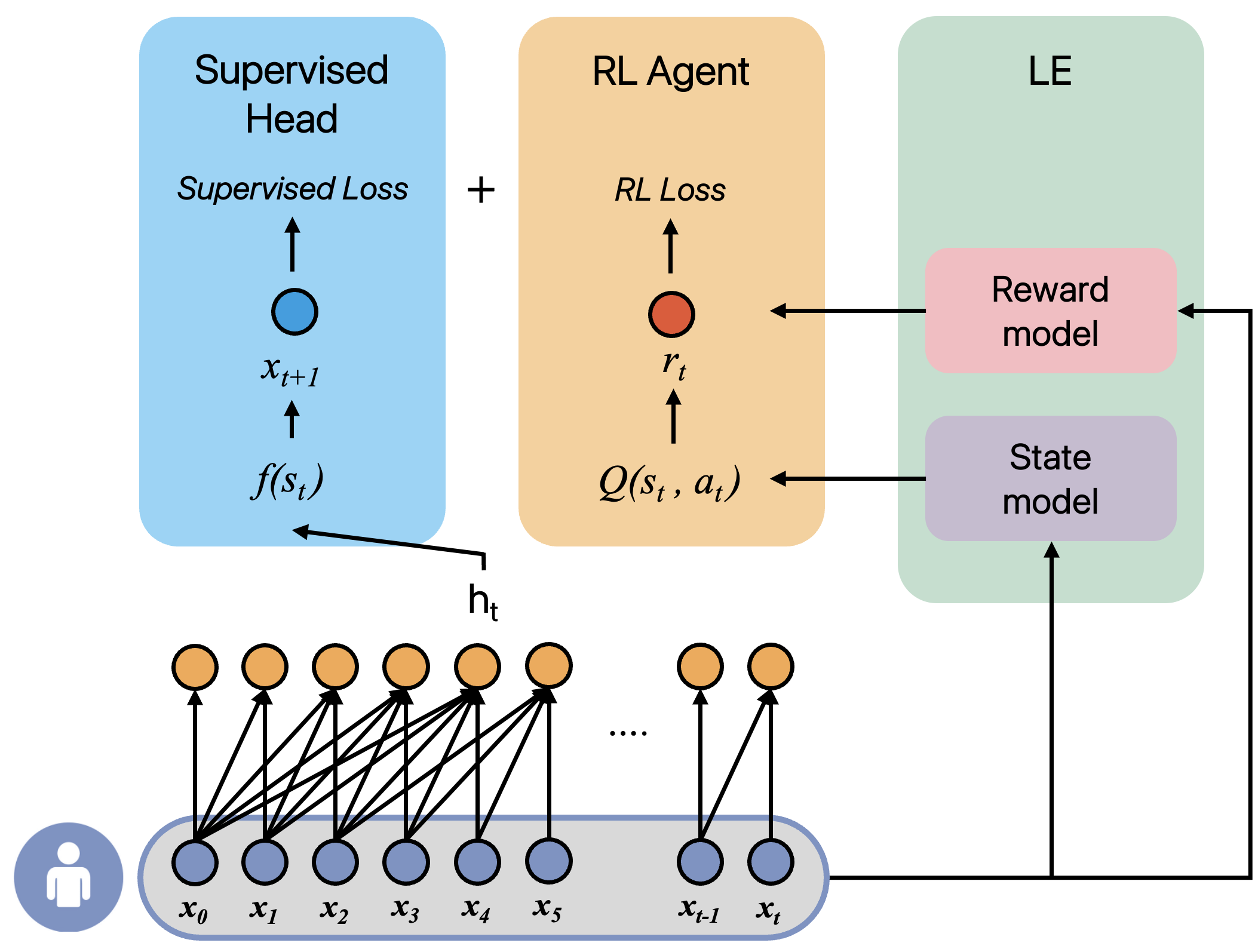}
        \label{figure1_b}
        }
    \end{minipage}
    \vspace{-0.5em}
    \caption{Self-supervised Reinforcement Learning for Recommendation (SSRL4R). (a) shows the previous offline structure, where the state for the RL agent is the hidden state from the sequential model, and the reward value is a predefined scalar. (b) shows our proposed structure, where the state is generated from a separate state model, and the reward is from a reward model.
    }
    % }\vspace{-0.2cm}
    \label{figure1}
\end{figure*}

Let $I$ denote the set of items in the system, $x_{1:t} = \{x_{1}, \ldots, x_{t}\}$ denote the user-item interaction sequence, where $x_{i} \in I(0<i \leq t)$ is the index of the interacted item ordered by timestamp. 
The goal is to recommend the next item $x_{t+1}$ at timestamp $t+1$ for the user from the whole item set $I$, such that the user might be interested in given the previous interactions $x_{1:t}$. Sequential recommendation methods~\cite{xin2020self} have been proposed to tackle the task as a self-supervised learning problem and map the sequence $x_{1:t}$ into the hidden state $h_t$ by a sequential model $G(\cdot)$, i.e., $h_t = G(x_{1:t})$. A fully connected layer then follows to map $h_t$ to classification logits (ranking scores) on all candidate items at the next step. In RL-based recommendation, the task is further modeled as an MDP, where users are treated as the environment. 
As Figure~\ref{figure1_a} shows, the state from the environment is represented by the user's historical interactions, and an RL agent is additionally trained to interact with users by taking actions (i.e., recommending items) to maximize the cumulative rewards. 
 
In previous
methods, the user state is taken (re-used) from the hidden state $h_t$ of the sequential model $G(\cdot)$, and the observed rewards are usually predefined w.r.t specific behaviours,
e.g., purchase=1, click=0.2. 
However, the uniform reward setting for all positive or negative actions fails to accurately capture the nuances of user preferences, resulting in an agent that is unable to discern the latent differences between items for a user. Moreover, user states generated via \(G(\cdot)\), based on implicit data comprised of item IDs, cannot reflect user behaviours on specific content.
In offline training, the agent is usually trained on fixed historical user-item interactions without probing the environments due to the significant costs associated with error-prone explorations, which introduces the challenge of insufficient positive and negative signals.
The proposed solution involves developing a state model to represent nuanced user states based on historical data, alongside with formulating an accurate reward model that assigns rewards contingent upon specific user states and actions. Additionally, we attempt to distill user's potential interests from the environment to augment the positive feedback for offline training, thereby reinforcing the exploration capabilities of RL-based RS.

\subsection{Large Language Model as Environment}
\label{method_le}

LLMs are considered as emergent world representations~\cite{Gurnee2024language,li2023emergent} with transfer learning capabilities. They can adapt to specific downstream tasks with minimal data through fine-tuning or prompt-based methods, e.g., optimization based on prompts and adapters that adjust a small number of parameters~\cite{houlsby2019parameter,he2022sparseadapter} to generate question answers~\cite{hu2023llm}. In this work, we leverage LLM as an offline user environment (LE) to return user feedback for RL-based RS attributes for three reasons. First, LLMs are ideal for creating environments that can simulate user queries/behaviours and feedback due to their ability to understand and generate natural language~\cite{he2023large,li2023prompt}. 
Second, environments created with LLMs significantly decrease the cost incurred by RL experiments in online settings, as error explorations could impair the user experience. Third, LLMs can generate feedback and shape rewards on various objectives, aiding RL-agents in learning complex signals that reflect user satisfaction. 
Therefore, we construct the LE based on the Transformer decoder-based LLM and use the latest Mistral 7B model 
for its simplicity and efficiency.

In summary, our objective is to efficiently tune LLMs with a small subset of offline user data, transforming them into a user environment (LE) capable of providing user feedback, including states, rewards, and predicted positive actions. 

\noindent\textbf{\textbullet{}Item tokenization.}
We first propose an initial step of tokenizing items based on their textual content, a preparation step for efficiently fine-tuning the LLM with user-item interaction data. Item tokenization addresses two issues.
First, when representing items by textual content,  e.g., descriptions of a product in Amazon, excessively long user-item interactions can lead to truncation and inefficiency~\cite{hua2023index}, and the dilution of user signals for training an LE.
Second, recent studies~\cite{rajput2023recommender} represent semantic items via outputs of an encoder fed with textual content to support another recommendation model. However, such item embeddings are not suitable as the input of the LLM since they deviate from the LLM's token embedding distribution.
Therefore, we aim to tokenize each item into the LLM's token embedding space $\mathcal W$ without losing its semantics and improve efficiency in learning an LE.

\begin{figure}[!t]
    \captionsetup[subfloat]%{captionskip=-5pt,nearskip=0pt,farskip=0pt}
    {}
    \centering
    \begin{minipage}[l]{0.22\textwidth}
    \centering
        \vspace{.75em}
        \subfloat[Item Tokenization]{%
        \includegraphics[trim={0 0 0 0}, clip, scale=0.11]{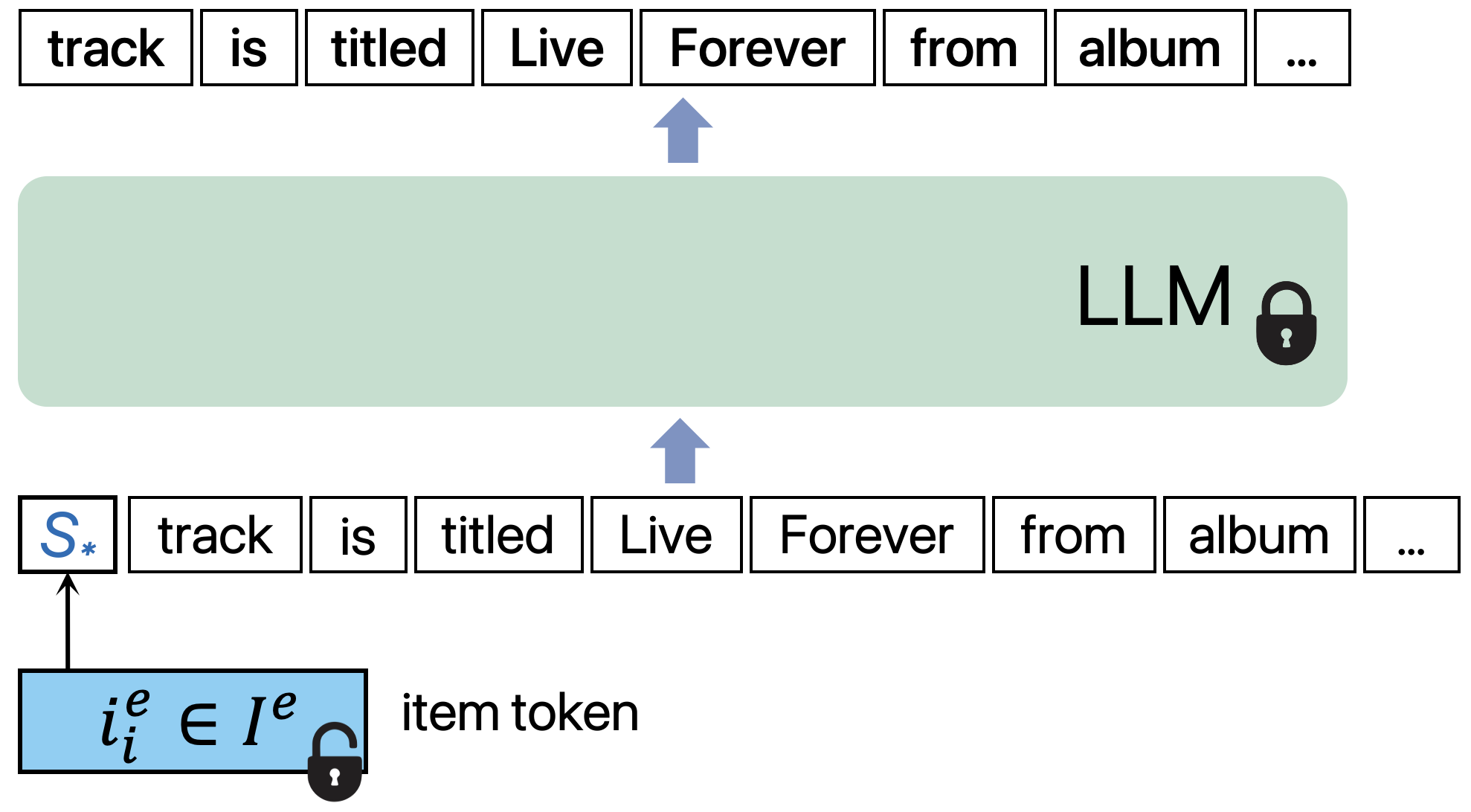}
        \label{figure2_a}
        }
    \end{minipage}
    \hspace{0.75mm}
    \begin{minipage}[r]{0.23\textwidth}
    \centering
        \subfloat[Prompt-based fine-tuning of LE]{%
        \includegraphics[trim={0 0 0 0}, clip, scale=0.11]{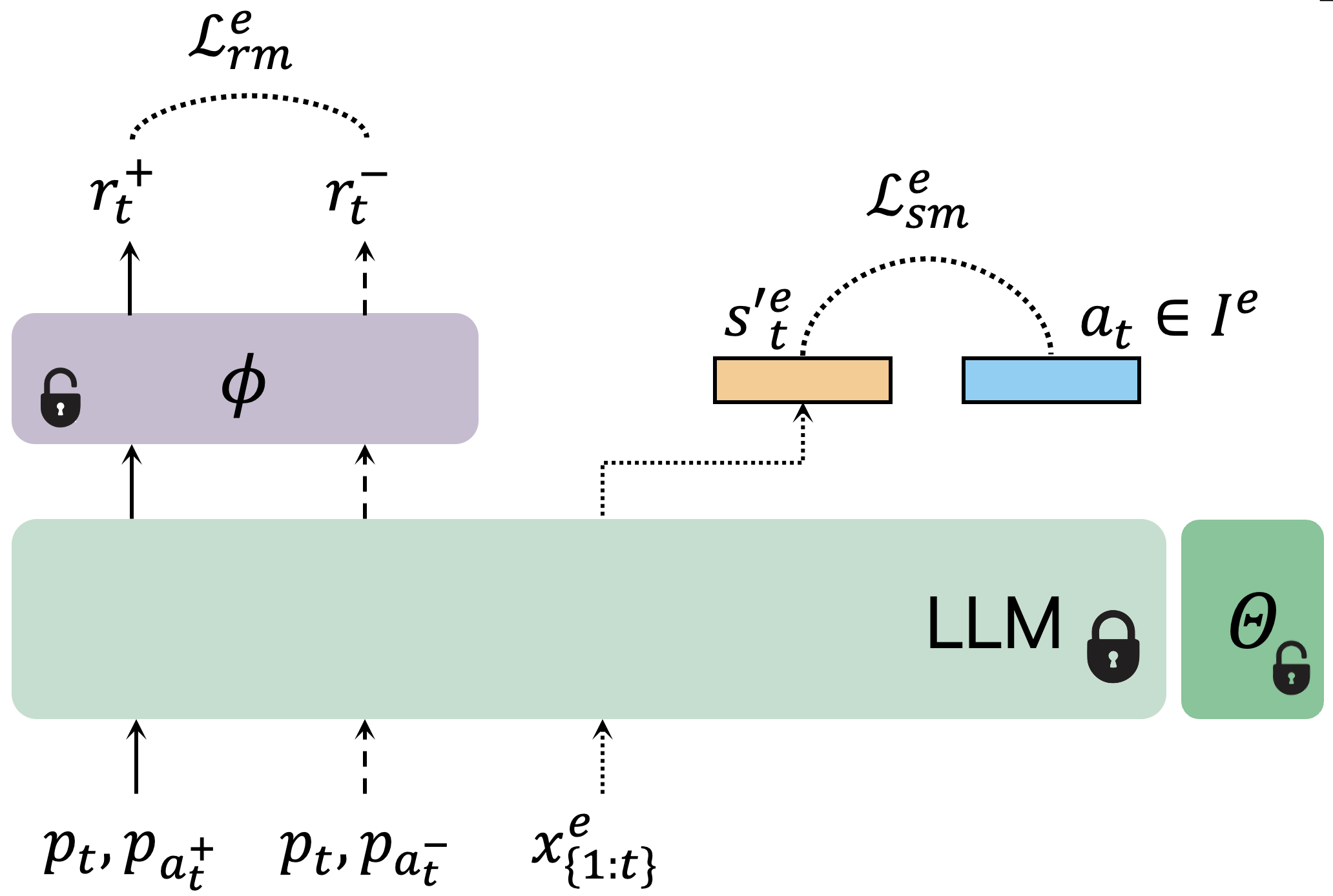}
        \label{figure2_b}
        }
    \end{minipage}
    \vspace{-0.5em}
    \caption{Our approach of adapting decoder-only LLM as Environment (LE). (a) we produce token $i^e_i \in I^e$ for item $i_i\in I$ by optimizing the objective of generating the next tokens of its textual content autoregressively. (b) we learn the LE by parameter-efficient adapters $\phi$ on a small subset of user data. 
    User-item token interactions $x^e_{1:t}$, where $x^e_i\in I^e$, is the input to generate the state representation $s^e_t$. We enhance the state representation by comparing the similarity between the state and actions through loss $\mathcal{L}_{sm}$.
    Reward prompt $p_t, p_{a_t}$ contains $x^e_{1:t}$ and action $a_t \in [a^+_t, a^-_t]$, where  $a^+_t$ is the positive action (next interacted item), and $a^-_t$ is the negative action (sampled uninteracted item). The action-specific reward for a user is produced by a score head $\theta$, and the LLM is trained by comparing user preferences for actions via loss $ \mathcal{L}_{rm}$.
     }
    % }\vspace{-0.2cm}
    \label{figure2}
\end{figure}

We achieve this goal by condensing the textual information of an item into a new item embedding space $I^e$ belonging to $\mathcal W$.
We refer to  
$i^e_i \in I^e$ as an item token, since it is not associated with an actual word, but is rather another representation of the items in $\mathcal W$.
Given an item $i_i \in I$ with textual content, we build a sentence $T$ and adopt an optimization-based approach that performs tokenization by updating the randomly initialized item token $i^e_i$ for each item:  
\\[1mm]
\fbox{
\parbox{0.96\columnwidth}{
%\hspace{-2mm}
\small\textbf{An example of $T$:} $S_*$ track is titled Live Forever from album Definitely Maybe, its artist is Oasis.
%\hspace{-2mm}
\vspace{-0.7mm}
}}\\[1.mm]
where $S_*$ signifies a placeholder of $i^e_i$.
$T$ is fed to
the decoder-only LLM. Figure~\ref{figure2_a} shows the process. The objective is to generate the next tokens of the sentence autoregressively, conditioned on the only-optimized item token. The final obtained item token encapsulates signals of descriptive content and serves as a semantic representation in LLM.
We finally obtain the environment item token set $I^e$ and the corresponding item ID embedding set $I$ in the system, where their indices are aligned.
The user-item interactions $x_{1:t}$ in the environment is denoted as $x^e_{1:t}$,where $x^e_{i} \in I^e(0<i\leq t)$.

\noindent\textbf{\textbullet{}Reward Model (RM).} To learn a reward model from a small amount of historical data, we fine-tune the LLM to take input as a reward prompt concatenated by a user prompt $p_t$ %interactions 
and an action prompt $p_{a_t}$
to output a scalar reward:\\[1.5mm]
\fbox{
\parbox{0.96\columnwidth}{
\small \textbf{Reward Prompt} is the concatenated text of $p_t$ and $p_{a_t}$: "$p_t p_{a_t}$", where\\
\small$p_t\gets$The user has listened to these tracks in chronological order: $x^e_{1:t}$\\
%, x^e_2,x^e_3$ \\
\small$p_{a_t}\gets$Compute the likelihood that $\mathit{item_X}$ be the next track to be listened to based on the listening history.
\vspace{-0.7mm}
}}\\[1.5mm]
where $\mathit{item_X} \in \{a^+_t, a^-_t\}$.   
$a^+_t$ is the truly interacted item $x^e_{t+1}$ at next step and $a^-_t$ is the negatively sampled one. As Figure~\ref{figure2_b} shows, the reward prompt $p_t,p_{a_t}$ guides the LLM to generate the reward score based on the similarity between user-item token interaction and the specific action. 
A linear network $\phi $ is added as score head to generate rewards by the last hidden state.
We use the Low-Rank Adaptation architecture (LoRA)~\cite{hu2021lora} adapter $\phi$ for each Transformer block for computational efficiency\footnote{We utilize the established prompt engineering~\cite{li2023prompt,liu2023pre} and adapter techniques~\cite{hu2021lora,bao2023tallrec}.
\label{footnote:pt_adpt}
}.  The loss function for updating the RM is defined as:

\begin{equation}
\label{loss_L_rm}
\mathcal{L}^e_{rm}= -log[ \sigma(r^e_{\theta + \phi}(p_{t}, p_{a^+_t})-  r^e_{\theta + \phi}(p_{t}, p_{a^-_t}))] ,
\end{equation}
where $ r^e\in [r^+_t, r^-_t]$  
is the scalar reward conditioned on the reward prompt
for action $a^+_t$ and $a^-_t$ at timestamp $t$, while $\theta$ and $\phi$ represent the trained parameters of the score head and the adapter.

\noindent\textbf{\textbullet{}State Model (SM).}
We learn an SM to refine state representations from historical interactions. 
As shown in Figure~\ref{figure2_b}, the user-item token interactions $x^e_{1:t}$ is fed to the LLM to generate the user state representation $\mathrm{~s'}^e_t$ at timestamp $t$, which is represented by the last hidden state of the outputs. 
Given that the state is modeled from a sequence of actions, where the difference between two consecutive states is the addition of the next interacted item in the input of the LE, there is a possibility that consecutive states could be quite similar. To ensure the model can differentiate states even when their interaction patterns are similar, we employ a contrastive loss: 
\begin{equation}
\label{loss_L_sm}
\mathcal{L}^e_{sm}= -\frac{1}{B} \sum_{j=1}^B log(\sigma(\mathrm{~s'}^e_t \cdot a^+_t-  \mathrm{~s'}^e_t \cdot a^j)) .
\end{equation}
where B is the batch size. For each sample,
$\mathrm{~s'}^e_t$ 
is the state at timestamp $t$ generated by 
$x^e_{1:t}$, 
$a^+_t$ is the next interacted item and $a_j$ is the $j$-th action of the batch.
Parameter $\phi$ is updated.

\noindent\textbf{\textbullet{}Discussion.}
The SM loss computes the disparity between the user state and token embeddings of true and sampled actions. While the RM loss takes both user interaction and sampled actions as input, the score differences are generated by the reward values. 
We train RM and SM simultaneously using a shared adapter on a small amount of offline historical user data, e.g., 10\% interactions (Figure~\ref{figure2_b}). 
After fine-tuning RM and SM, we obtain the LE that can generate user feedback for our RL-based recommendation tasks. 

\subsection{LE for RL-based Recommenders}
\label{method_le_rl}
\begin{figure}[!t]
      \centering
        \includegraphics[trim={0 0 0 0}, clip, scale=0.19]{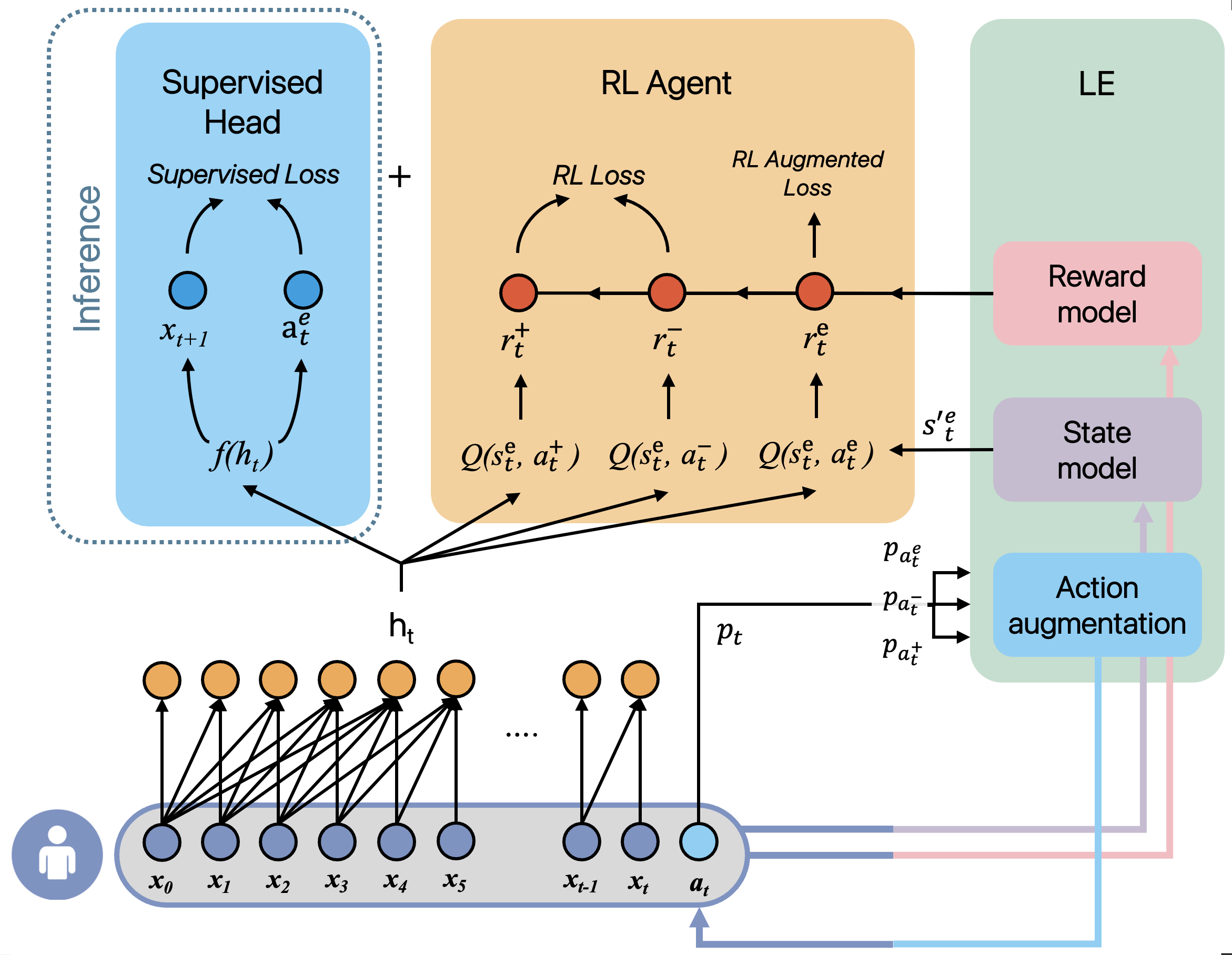}
    \vspace{-0.5em}
    \caption{
    Structure of LEA. \textit{Left}: the LE is applied to offline data. $(x_1^{(e)}, x_2^{(e)}, ... ,x_t^{(e)}) $ denotes the user-item interaction $x_{1:t} $ for the sequential model, where $x_i \in I$, and $x_{1:t}^e$ denotes the user-item token interaction for the LE, where $x^e_i \in I^e$. $a_t^e$ is the positive action predicted by LE. \textit{Right}: RL policy is trained via the original Q-loss and the augmented one $\mathcal{L}_{aq}$; the base sequential model is jointly trained through RL loss and the supervised loss over the original next item and the augmented one $\mathcal{L}_{ah}$ over $a_t^e$.
    }
    \label{figure3}
\end{figure}

We reframe the RL-based sequential recommendation as a novel LE-based Markov Decision Process (LEMDP), discussed in section~\ref{task_formulation}. The process by which the agent interacts with the LE to obtain user feedback is shown in Figure~\ref{figure3},  which can be represented by a tuple of $(\mathcal{S}^e, \mathcal{A}, \mathcal{P}, r^e, \gamma )$:

\noindent\textbf{\textbullet{} State space $\mathcal{S}^e$ generated by the State Model (SM) $\mathcal{E}^e$ in LE.} The set of states with the time series, modeled by the user's historical interacted item tokens $x^e_{1:t}$ via the SM and the hidden state $h_t$ from the sequential model:% i.e., $\mathrm{~s}^e_t = \mathcal{E}^e(x^e_{1:t})$.

\begin{gather}
\mathrm{~s'}^e_t = \mathcal{E}^e(x^e_{1:t}), \label{le_s'} \\
\mathrm{~s}^e_t = {s'}^e_t \parallel h_t, \text{where } h_t=G(x_{1:t}) \label{le_s}
\end{gather}

\noindent\textbf{\textbullet{}Action space $\mathcal{A}$.} The discrete action set is comprised of the candidate items. Taking action in LEMDP means recommending items. In the offline data, the action $a_t\in A$ at timestamp $t$ is the interacted item in the next step, i.e., $a_{t}=x^e_{t+1}$.
     
\noindent\textbf{\textbullet{}Reward Model (RM) $r^e$ in LE.}
     The reward function that returns immediate reward $r_t$ as the state $\mathrm{~s}^e_t$ and the action $a_{t}$ taken by the agent at step $t$ are observed: 
\begin{gather}
  r_t = {r}(s^e_t, a_t) = r^e(p_{t}, p_{a_t})
\end{gather}
where $p_{t}, p_{a_t}$ is the reward prompt.

\noindent\textbf{\textbullet{}State Transition Function $\mathcal{P}$.}
     The transition function describes the next state from the environment given the observed action and the current state. When learning from offline data, only the positive actions affect the state.
     
\noindent\textbf{\textbullet{} Discount factor $\gamma $.}  This defines the discount factor to the future rewards, where $\gamma  \in [0,1]$.

RL aims to learn a target policy $\pi_\psi(a_t | s^e)$ that maps the state  $s_e\in\mathcal{S}^e$ to an action distribution $a\in\mathcal{A}$ by maximizing the expected cumulative rewards (returns), where $\psi$ denotes the parameters:
\begin{equation}
\max _{\pi_\psi} \mathbb{E}_{\tau \sim \pi_\psi}[R(\tau)] \text {, where } R(\tau)=\sum_{t=0}^{|\tau|} \gamma^t r\left(\mathrm{~s}^e_t, a_t\right) \text {, }
\end{equation}
where $\tau$ denotes the trajectory of $(s^e_t, a_t, s^e_{t+1})$. We apply the LE to the value-based Q-learning algorithm~\cite{xin2022supervised} to train the target policy. 

Following the objective in section~\ref{exp_rq}, we improve 
the SNQN and SA2C~\cite{xin2022supervised} frameworks to combine supervised learning of the sequential model $G$ and RL via the agent-LE interactions. More specifically, given the input item sequence $x_{1: t}$ and 
$G(\cdot)$ for self-supervised learning, the hidden representation is formulated as ${h}_t=G\left(x_{1: t}\right)$. Then, $h_t$ is used into a fully connected layer and a softmax function to output ranking scores over the candidate items:
\begin{gather}
\label{user_state}
\hat{Y}_i = softmax(\delta (W_{u}{h_t}+b_{u})),
\end{gather}
where  $\hat{Y}_i  = [ \hat{y}_1,  \hat{y}_2, ...,  \hat{y}_n ] \in \mathbb{R}^{n}$. 
 $n$ is the number of items,
$\delta$ is the activation function, $W_{u} \in \mathbb{R}^{d\times n}$ are trainable parameters, and $b_{u} \in \mathbb{R}^{n}$ is the bias vector. The self-supervised part is trained via the cross-entropy
loss:
\begin{equation}
\label{loss_L_h}
\mathcal{L}_{h}= - \sum_{i=1}^n y_i \log \left(\hat{y}_i\right),
\end{equation}
$y_{i}= 1$ if the user interacted with the $i$-th item in the next timestamp. Otherwise, $y_{i}= 0$.

Regarding the Q-learning network, we obtain the state $s^e_t$ by the SM in LE and compute the Q-value as follows:
\begin{gather}
    Q(s^e_t, a_t) = \delta (W_{q}{s^e_t}+b_{q})
\end{gather}
where $\delta$ is the activation function, $W_{q} \in \mathbb{R}^{d\times n}$ are trainable parameters, and $b_{q} \in \mathbb{R}^{n}$ is the bias vector. The one-step time difference (TD) Q-learning loss to improve SNQN is defined as:
\begin{gather}
\label{loss_L_q}
\mathcal{L}_q=\underbrace{(r^e(p_t, p_{a_t^+})+\gamma \max _{a^{\prime}} Q(\mathrm{~s}^e_{t+1}, a^{\prime})-Q(\mathrm{~s}^e_t, a_t^{+}))^2}_{L_p: \text {~positive TD error}} \\
+\underbrace{(r^e(p_t, p_{a_t^-})+\gamma \max _{a^{\prime}} Q(\mathrm{~s}^e_t, a^{\prime})-Q(\mathrm{~s}^e_t, a_t^{-}))^2}_{L_n: \text {~negative TD error}},
\end{gather}
where $a_t^{+}$and $a_t^{-}$are the positive action and sampled unobserved (negative) action at timestamp $t$, respectively. 
In our method, we only sample one negative action.
We follow the assumption that taking negative actions will not affect the state. Therefore, 
the maximum operation is performed in $Q\left(\mathrm{~s^e}_t, a^{\prime}\right)$ other than $Q(\mathrm{~s}^e_{t+1}, a^{\prime})$ in the negative TD error $L_n$. 
In SA2C, the advantage Q-value~\cite{xin2022rethinking} is calculated to formulate the supervised loss:
\begin{gather}
    A(\mathrm{~s}^e_t, a_t^{+})=(Q(\mathrm{~s}^e_{t}, a^{+})-Q(\mathrm{~s}^e_t, a_t^{-}))/2
\end{gather}
Then the supervised loss $\mathcal{L}_h$ is formulated as  $\mathcal{L}_h \gets \mathcal{L}_h \cdot  A(\mathrm{~s}^e_t, a_t^{+})$ for SA2C. The corresponding loss function used in the LE to directly improve the above RL frameworks is formulated as follows:
\begin{gather}
\label{loss_L_c}
    \mathcal{L}_c = \mathcal{L}_h+\mathcal{L}_q
\end{gather}

\begin{algorithm}[!b]
 \caption{Training procedure of LEA}
 \label{algLEA}
 	\begin{algorithmic}[1]
        \renewcommand{\algorithmicrequire}{\textbf{Input:}} 
        \renewcommand{\algorithmicensure}{\textbf{Output:}}
        \Require
        user-item interaction sequence set $\mathcal{X}$, recommendation model $G$, reinforcement head $Q$, supervised head, environment LE comprised of state model $\mathcal{E}^e$  and reward model $r^e$, 
        item set $I$,
        item token set $I^e$
        \Ensure
        all parameters in the learning space $\Theta$
        \State Initialize all trainable parameters
        \State Create $G'$ and $Q'$ as copies of $G$ and $Q$, respectively
        \Repeat 
            \State Draw a mini-batch of $(x_{1:t},a_t)$ from $\mathcal{X}$, the corresponding user-item token interaction $x^e_{1:t}$, sample a negative action $a^-_t$, generate augmented action $a^e_t$
            %\State $r_t = SM(p_t,p_{a_t})$,
            \State ${s^e}_t=G(x_{1:t}) || \mathcal{E}^e(x^e_{1:t})$, ${s^e}'_t=G'(x_{1:t}) || \mathcal{E}^e(x^e_{1:t})$ 
            \State ${s^e}_{t+1}=G(x_{1:t+1}) || \mathcal{E}^e(x^e_{1:t+1})$, ${s^e}'_t=G'(x_{1:t+1}) || \mathcal{E}^e(x^e_{1:t+1})$ 
            %\State 
            \State Generate random variable $z \in (0,1)$ uniformly
            \If {$z\leq0.5$}
                \State $a^*=\text{argmax}_a\text{ }Q({s^e}_{t+1},a)$
                %\State $L_q=(r+\gamma Q'(\mathbf{s^e}'_{t+1},a^*)-Q(\mathbf{s^e}_t,a_t))^2$,        $L_{aq}=(r+\gamma Q'(\mathbf{s^e}'_{t+1},a^*)-Q(\mathbf{s^e}_t,a^e_t))^2$
                \State Calculate $\mathcal{L}_{q}$ and $\mathcal{L}_{aq}$ according to Eq.(\ref{loss_L_q}) and Eq.(\ref{loss_L_aq})                
                \State Calculate $\mathcal{L}_{h}$ and $\mathcal{L}_{ah}$ according to Eq.(\ref{loss_L_h}) and Eq.(\ref{loss_L_ah})
                \State Perform updates by $\nabla_\Theta \mathcal{L}$, i.e., Eq.~\ref{loss_L}
            \Else
                \State $a^*=\text{argmax}_a\text{ }Q'({s^e}'_{t+1},a)$
                %\State $L_q=(r+\gamma Q(\mathbf{s^e}_{t+1},a^*)-Q'(\mathbf{s^e}'_t,a_t))^2$
                \State Calculate $\mathcal{L}_{q}$ and $\mathcal{L}_{aq}$ according to Eq.(\ref{loss_L_q}) and Eq.(\ref{loss_L_aq})                
                \State Calculate $\mathcal{L}_{h}$ and $\mathcal{L}_{ah}$ according to Eq.(\ref{loss_L_h}) and Eq.(\ref{loss_L_ah})
                \State Perform updates by $\nabla_\Theta \mathcal{L}$, i.e., Eq.~\ref{loss_L}
            \EndIf
        \Until converge
        \State return all parameters in $\Theta$
 	\end{algorithmic}
 \end{algorithm}
 
\subsection{Augmentation via LE}
\label{method_lea}

Since the recommender is trained on historical data without online exploration, both the supervised model and the RL agent can only be trained on user-item interactions that exist in the offline training data. As a result, the models may fail to estimate the value functions for unseen user feedback.
We propose an LE augmentation (LEA) 
method to augment the historical user data for offline training. We prompt the LE to predict items $a^{e}_t$  the user is likely to interact with in the next timestep, these predicted positive actions are used for both supervised learning and Q-learning. We build an augmentation prompt template constructed by the user prompt $p_t$ and a selection prompt $p_{l_t}$ to generate feedback:
\\[1.5mm]
\fbox{
\parbox{0.97\columnwidth}{
\small \textbf{Augmentation Prompt} is the text concatenation of $p_t$ and $p_{l_t}$: "$p_t p_{l_t}$".\\
\small$p_{l_t}\gets$In the list of following 5 tracks:[$list_t$], based on the history, select the number of the track that he is most likely to continue to listen to.
\vspace{-0.7mm}
}}\\[1.5mm]
where $list_t$ is the token sequence of the sampled items. After an early training stage, we sample the top-5 items from the supervised head to construct $list_t$.
We feed the augmentation prompt into the LE and obtain the predefined item position classification (e.g., ``\texttt{first}'' for the first item in the list). Then we select the item with the highest label score as the predicted positive action.
The predicted item will be used to augment both the supervised learning for the sequential model and the Q-learning for RL agent:
\begin{equation}
\label{loss_L_ah}
\mathcal{L}_{ah}= - \sum_{i=1}^n y^e_j \log \left(\hat{y}_j\right),
\end{equation}
\begin{gather}
\label{loss_L_aq}
\mathcal{L}_{aq} = (r^e(p_t, a_t^{e})+\gamma \max _{a^{\prime}} Q(\mathrm{~s}^e_{t+1}, a^{\prime})-Q(\mathrm{~s}^e_t, a_t^{e}))^2
\end{gather}
$y^e_{j}= 1$ if the LE predicts that the user will interact with the $j$-th item at the next timestamp. Otherwise, $y^e_{j}= 0$. 
The max operation of $\mathcal{L}_{aq}$ is performed in  
$Q(s^e_{t+1}, a^{\prime})$, since the positive actions will transit the current state to the next state.

\noindent\textbf{\textbullet{}Training.} We jointly train the supervised and Q-learning loss on the original offline user data, with the augmented supervised and Q-learning loss on the predicted positive feedback:
\begin{gather}
\label{loss_L}
\mathcal{L} =  \mathcal{L}_c +  w_{ah} \mathcal{L}_{ah} + w_{aq}\mathcal{L}_{aq} %=  L_h + L_q +  w_a(L_{ah} + L_{aq})
\end{gather}
where $w_{ah}$ and $w_{aq}$ are the weights of augmented supervised learning loss and Q-learning loss, respectively.

\begin{table}[!t] %\footnotesize
\fontsize{9}{10}\selectfont
\setlength{\tabcolsep}{2.5pt}
%\vspace{-0.3em}
\caption{Dataset statistics. seq. denotes sequence, inter. denotes interaction.
}
\label{dataset_statistics}
\vspace*{-4mm}
\centering
\begin{tabular}{lcccl}

\toprule
Dataset    & \#item          & \#seq.  &\#inter.   &Item content       \\ 
\midrule
LFM      & 18,297         &11,073     &146,255 &\small title; album; artist \\
Industry   &5,814         &10,935   &71,872 & \small title; category; brand; description \\
\bottomrule
\end{tabular}
\vspace{-0.4cm}
% \end{wraptable}
\end{table}

\subsection{Discussion}
\label{method_discussion}
Our goal is to distill knowledge from LLMs into the RL-based RS. Previous methods train LLMs to act as recommenders, which are difficult to deploy in real world settings due to low inference speed. During the inference stage, we only use the sequential model with the supervised head to generate the top-$k$ recommendations without compromising efficiency.
We compare LEA with two SOTA RL frameworks: SNQN and SA2C~\cite{xin2022supervised}. 
The difference between LEA and the direct application of LE in existing RL frameworks is that it implements an augmentation method through the LE. We illustrate the training procedure of LEA utilizing LE as both state and reward models in Algorithm~\ref{algLEA}, where double Q-learning~\cite{hasselt2010double} is adopted to alternatively train two copies of trainable Q-networks.
The function of LE as a state model or a reward model will be discussed in the experimental section by directly replacing the corresponding part of existing RL methods.

\section{Experimental setup}
\subsection{Research Questions} 
\label{exp_rq}

We detail the experimental setup for validation of our augmentation method via the LLM-based environment (LEA).
We aim to answer the following research questions:
\\
\textbf{RQ1:} How does LEA, integrated with feedback from LE, perform compared with the existing RL-based 
recommendation frameworks?
\\
\textbf{RQ2:} Does action augmentation affect the performance of LEA?
\\
\textbf{RQ3:} How does the user feedback from LE, i.e., rewards and state, affect the RL-based recommenders?
\\
\textbf{RQ4:} Do various fine-tuning strategies for LE affect its performance, i.e., item tokenization, data scale, state model loss, the use of LLMs?

\subsection{Datasets} 
\label{Datasets}

We perform experiments on two real-world datasets  (Table~\ref{dataset_statistics}): \textbf{LFM}~\cite{schedl2016lfm} and \textbf{Industry}~\cite{ni2019justifying}. The LFM dataset was collected from the music streaming platform Last.fm\footnote{http://www.last.fm/api} 
and contains more than one billion listening events by 120k users. 
We sample a subset of 3-day listening events on tracks as experimental data.
The textual description for each track (item) includes its title, album and the artist. We filter out sequences with fewer than three interactions and tracks listened to less than three times, obtaining a dataset comprising 18,297 items and 11,073 user sequences, with a total of 146,255 interactions.
The second Industry dataset is from the publicly available Amazon review dataset\footnote{https://nijianmo.github.io/amazon} of the Industrial and Scientific category. The textual description for each product (item) includes the title, category, and product description. Similarly, we filter out sequences with fewer than three interactions and products reviewed less than three times, yielding a dataset containing 5,814 items and 10,93 sequences, with 71,872 interactions.

\subsection{Baselines}
We compare LEA with two state-of-the-art 
RL frameworks: \textbf{SNQN} and \textbf{SA2C}~\cite{xin2020self,xin2022supervised} which are introduced in section~\ref{method_le_rl}, under two backbone sequential models: \textbf{GRU4Rec}~\cite{hidasi2015session}, the first sequential recommendation RS based on RNN, and \textbf{SASRec}~\cite{kang2018self}, a renowned sequential recommendation model based on self-attention. \textbf{Normal} denotes the original sequential models with normal supervised loss.
By applying LEA method, we obtain the following strategies: (1) \textbf{LEAR} denotes training RL framework with the rewards from LE. (2) \textbf{LEAS} denotes that the state for the RL component is derived from LE. (3) \textbf{LEASR} denotes that both state and rewards are from the LE. In SA2C, the advantage calculation allows the training of the RL policy to affect the updates of the sequential model when only ${s'}^e_t$ is used, therefore, we introduce an additional method (4) \textbf{LEAS$'$R} for SA2C that the state for RL agent is solely from LE, i.e., $\mathrm{~s}^e_t = {s'}^e_t$ in Eq.~\ref{le_s'}. 
In ablation studies,  we replace each feedback in the compared RL framework with the corresponding one from LE to examine the LE environment: (1) \textbf{LER} and (2) \textbf{LES} denote the rewards and states in the baseline RL methods are generated by LE, respectively. Moreover, (3) \textbf{LEA} denotes only the augmented training strategy applied to the baseline models.

\subsection{Metrics} 
We apply two commonly used metrics: 
Hit Ratio (HR$@k$)~\cite{xin2022supervised} and Normalized Discounted Cumulative Gain (NDCG$@k$)~\cite{jarvelin2002cumulated} to measure the relevance of recommended items for the evaluated sessions, where $k \in \{5, 10, 20\}$.
HR$@k$ evaluates whether the ground-truth item is in the top-$k$ positions
of the recommendation list. NDCG$@k$, $k \in \{5, 10, 20\}$ measures the rank of the ground-truth item in the top-$k$ recommendation list. 
We report the average results over all interactions of the test sequences. 

\subsection{Implementation Details} 
For \textit{training the LE}, we perform 300 iterations with a learning rate of $5e^{-3}$ for optimization-based item tokenization.
The sequence length for the fine-tuning of LE for state generation is set to 10. The epochs for all datasets is 10 with a batch size of 20. We employ the Adam optimizer for fine-tuning. The batch size and epochs is 10 for all datasets. For efficient training, we only use a 10\% subset of the original dataset to obtain the LE. For all \textit{compared models and our LEA methods}, the length of sequences is set to 10, and a padding token is added to shorter sequences for all datasets. 
We train all models and variants with the Adam optimizer~\cite{xin2020self,stamenkovic2022choosing}. The mini-batch size is 100 for LFM and Industry. The learning rates are $1e^{-3}$ for LFM, and $2e^{-3}$ for Industry. We evaluate the validation set every 1,000 and 500 steps of updates on LFM and Industry, respectively. All experiments are performed in one H100 GPU. To ensure a fair comparison, the item embedding size and hidden size are set to 64 for all models. Since there is only one kind of behaviour, we set the reward to 1 for all interactions in original SNQN and SA2C~\cite{xin2022supervised}.

\FloatBarrier
\begin{table*}[h!]
\centering
\caption{Performance on LFM and Industry datasets. NG is short for NDCG.
Boldface denotes the highest score and the second-best scores are marked with \underline{\quad}. $*$ denotes the significance $p$-value < 0.01 compared with second best baseline.  
}
\label{exp_main_results}
\vspace*{-4mm}
\begin{tabular}{lcp{0.85cm}<{\centering}p{0.85cm}<{\centering}p{0.85cm}<{\centering}p{0.85cm}<{\centering}p{0.85cm}<{\centering}p{0.85cm}<{\centering}cp{0.85cm}<{\centering}p{0.85cm}<{\centering}p{0.85cm}<{\centering}p{0.85cm}<{\centering}p{0.85cm}<{\centering}p{0.85cm}<{\centering}} 
\toprule
\multirow{2}{*}{Model} & &\multicolumn{6}{c}{LFM} &  &\multicolumn{6}{c}{Industry}\\
\cmidrule{3-8} \cmidrule{10-15}

 & & HR@5 & NG@5 & HR@10 & NG@10 & HR@20 & NG@20 & & HR@5 & NG@5 & HR@10 & NG@10 & HR@20 & NG@20         \\ 
%\cmidrule{}
\midrule
\multirow{11}{*}{SASRec}  &Normal &0.2798 	&0.2324 &0.3115	&0.2646	&0.3271	&0.2685 &&0.0788 &0.0684&0.1171	&0.0909	&0.1374	&0.0961\\
\cmidrule{2-15}
 &SNQN &0.3001 	&0.2602 &0.3337	&0.2828 &0.3459 &0.2855	&&0.0926 &0.0738&0.1202	&0.0913	&0.1419	&0.0968\\
\cdashline{2-15}
%\cmidrule{2-15}

&LEAR &0.3286  &0.2879 &0.3504	&0.2951   &0.3721	&0.3005   &	&0.1027&0.0864&0.1232	&0.0929	&0.1514	&0.0983\\
&LEAS &0.3298&0.2905&0.3529&0.2981&\textbf{0.3773}$^*$&0.3039&&0.1047&0.0856&\textbf{0.1281}$^*$&0.0931&\textbf{0.1581}$^*$&\textbf{0.1008}$^*$\\
&LEASR  &\textbf{0.3323}$^*$  &\textbf{0.2914}$^*$ &\textbf{0.3533}$^*$&\textbf{0.2982}$^*$&0.3772&\textbf{0.3042}$^*$&&\textbf{0.1059}$^*$&\textbf{0.0866}$^*$&0.1245&\textbf{0.0933}$^*$ &0.1538   &0.1001	\\
\cmidrule{2-15}

&SA2C &0.2902 	&0.2501 &0.3372	&0.2851	&0.3516	&0.2885 &&0.0931 &0.0735&0.1205	&0.0912	&0.1416	&0.0966\\
\cdashline{2-15}
%\cmidrule{2-15}

&LEAR &0.3269 	&0.2573  &0.3429	&0.2942   &0.3571	&0.2977   & &0.1037&0.0861&0.1231&0.0922&0.1525&0.0996	\\
&LEAS &0.3302&0.2907&0.3533&0.2982&0.3773&0.3043& &\textbf{0.1055}$^*$&0.0868&\textbf{0.1278}$^*$&0.0938&0.1528&0.1002\\
&LEAS$'$R &0.3298  &0.2853  &0.3511	&0.2931	&0.3706	&0.2981 &   &0.1036&\textbf{0.0874}$^*$&0.1264&\textbf{0.0948}$^*$&0.1521&\textbf{0.1012}$^*$ \\
&LEASR  &\textbf{0.3325}$^*$  &\textbf{0.2926}$^*$  &\textbf{0.3572}$^*$	&\textbf{0.3011}$^*$&\textbf{0.3801}$^*$	&\textbf{0.3072}$^*$ &   &0.1042&0.0862&0.1261&0.0932&\textbf{0.1533}$^*$&0.1001	\\

\bottomrule

\multirow{11}{*}{GRU4Rec} &Normal &0.2757&0.2536&0.2934&0.2594 &0.3113	&0.2641	&	&0.0901 &0.0751&0.1128 &0.0824&0.1399&0.0893\\
\cmidrule{2-15}
&SNQN &0.2781&0.2526&0.2931&0.2575&0.3103&0.2619&&0.0891&0.0727 &0.1083 &0.0896	&0.1392	&0.0931	\\
\cdashline{2-15}
%\cmidrule{2-15}

&LEAR  &0.3211&0.2878&0.3407&0.2941&0.3574&0.2983&&0.0972&0.0887&0.1205&0.0907&0.1466&0.0975  	\\
&LEAS &0.2881&0.2596&0.3134&0.2679&0.3359&0.2735&&0.0976&0.0826&0.1232&0.0909&0.1525&0.0982\\
&LEASR 
&\textbf{0.3234}$^*$&\textbf{0.2909}$^*$&\textbf{0.3464}$^*$&\textbf{0.2983}$^*$&\textbf{0.3667}$^*$&\textbf{0.3034}$^*$  &  &\textbf{0.0978}$^*$&\textbf{0.0892}$^*$&\textbf{0.1238}$^*$&\textbf{0.0915}$^*$ &\textbf{0.1557}$^*$&\textbf{0.0998}$^*$\\
\cmidrule{2-15}
&SA2C &0.2896&0.2743&0.3173&0.2797&0.3337&0.2838&&0.0865 &0.0705 &0.1145	&0.0842	&0.1398	&0.0913 \\
\cdashline{2-15}
%\cmidrule{2-15}

&LEAR  &0.3142  &0.2782 &0.3333	&0.2844   &0.3518	&0.2891   &&0.0954  &0.0816	&0.1191   &0.0883	&0.1421   &0.0948	\\
&LEAS &0.3139&0.2846&0.3364&0.2921&0.3588&0.2977&&0.0986&0.0814&0.1224&0.0891&0.1531&0.0968\\
&LEAS$'$R &0.3048&0.2795&0.3214&0.2849&0.3393&0.2894&&0.0959&0.0818&0.1195&0.0889&0.1416& 0.0953\\
&LEASR  &$\textbf{0.3281}^*$ &\textbf{0.2926}$^*$&\textbf{0.3502}$^*$&\textbf{0.2998}$^*$&\textbf{0.3711}$^*$&\textbf{0.3051}$^*$&  &\textbf{0.0981}$^*$	&\textbf{0.0828}$^*$	&\textbf{0.1234}$^*$	&\textbf{0.0892}$^*$ &\textbf{0.1559}$^*$   &\textbf{0.0969}$^*$	\\

\bottomrule
\end{tabular}
\end{table*}

\section{Experimental results}
\subsection{Performance Comparison (RQ1)}

The performance comparison of the two public datasets is shown in Table~\ref{exp_main_results}. We observe the following: 
(1) In contrast to standard sequential models, our RL-based methods consistently outperform across all datasets. This demonstrates that the integration of augmented RL and supervised learning, with feedback from LE, contributes to enhancing recommendation performance. (2) LEA further improves the performance over the RL frameworks of SNQN and SA2C. This suggests that the augmentation method improves the learning performance on the supervised component, as well as the RL head. (3) On the LFM dataset, LEASR achieves the highest results on both HR and NDCG metrics, across two frameworks, outperforming other strategies. This indicates that the state and reward obtained from LE, when combined, can significantly improve the performance of RL-based recommenders.
On the Industry dataset, the overall relative improvement of the RL method by the LEA approach is not as pronounced as on the LFM dataset. This could be attributed to the LLM potentially incorporating more knowledge about music, an artefact due to its exposure to such content during pre-training, compared to product review data from the Amazon dataset. In conclusion, the proposed LEA methods are effective to improve the recommendation performance of RL frameworks.

\subsection{Effect of Action Augmentation (RQ2)}
We assess the impact of the LE generated augmented positive actions. We first examine the influence of the feedback on supervised learning (sv) and Q-learning (q) separately. Here, sv+q denotes that both components are augmented. Results are reported on two datasets utilizing the SASRec as the backbone. As Table~\ref{exp_aug_results} shows, the first pair is the original sequential models (Normal) and its sv augmented version. We can see that when the augmented samples are taken as additional supervised labels, the recommendation performance increases in all cases and datasets. This confirms the LE's capability to generate high-quality positive samples for offline training. 
The second comparison is considering the SA2C framework and augmentation on sv, q, and sv+q. We observe that the sv augmentation produces the same trends as in the first pair we compare. Furthermore, the improved performance of q over SA2C indicates the effectiveness of augmenting the Q-loss with the positive feedback. Notably, when combining sv and q augmentations, sv+q on the Industry dataset achieves better performance than the individual methods, demonstrating the potential of joint augmentation. On the LFM dataset, sv+q  surpasses the baseline and other augmentation methods in most cases, which suggests that the predicted positive samples may require specific reward configurations, distinct from the truth action, to fully realize the benefit. 

\begin{figure}[!t]
    \captionsetup[subfloat]%{captionskip=-5pt,nearskip=0pt,farskip=0pt}
    {}
    \centering
    \begin{minipage}[t]{0.23\textwidth}
        \subfloat[]{%
        \includegraphics[trim={0 0 0 0}, clip, scale=0.22]{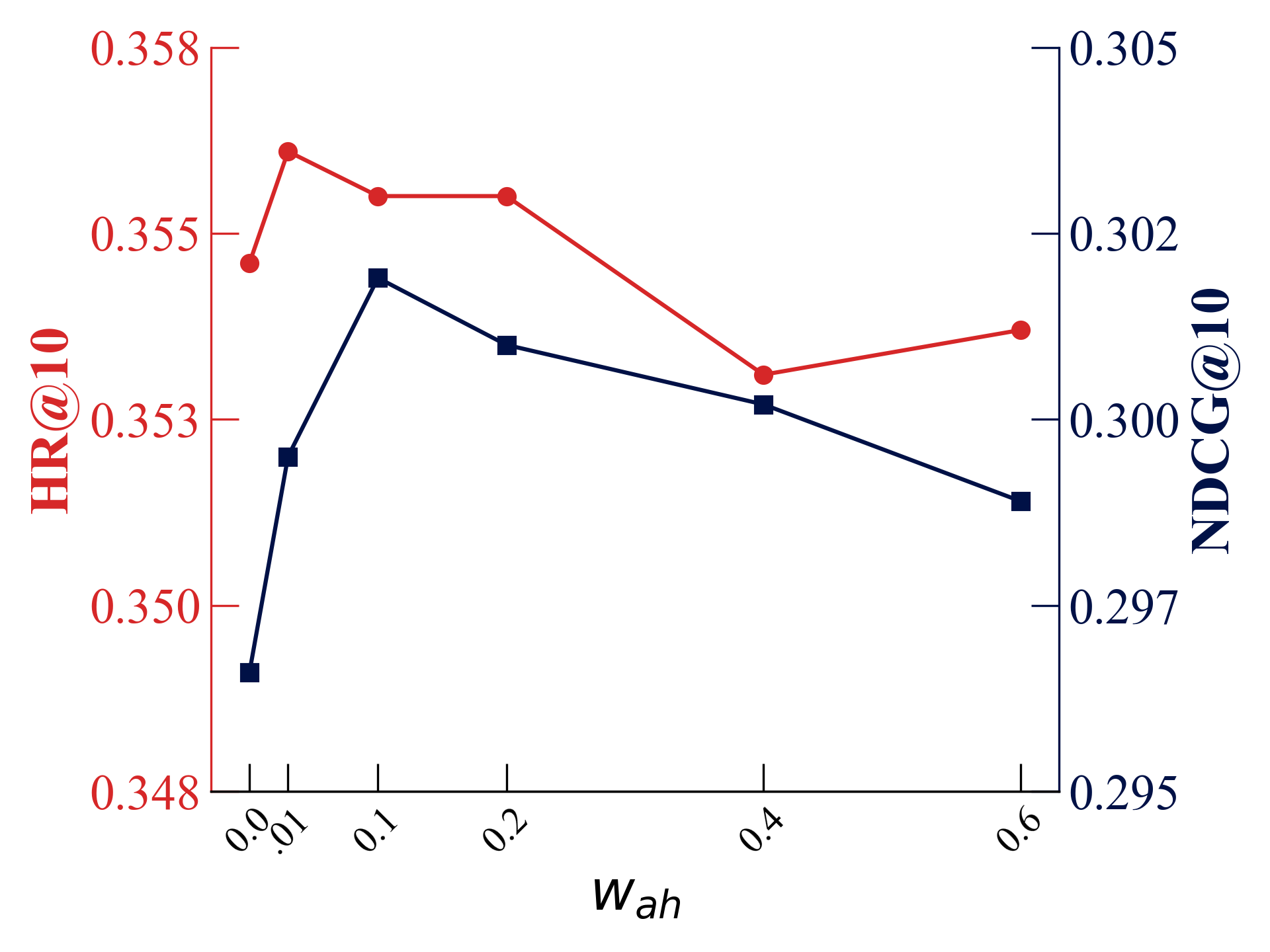}
        \label{effect_wah}
        }
    \end{minipage}
    \hspace{0.07mm}
    %\hfill
    \begin{minipage}[t]{0.23\textwidth}
        \subfloat[]{%
        \includegraphics[trim={0 0 0 0}, clip, scale=0.22]{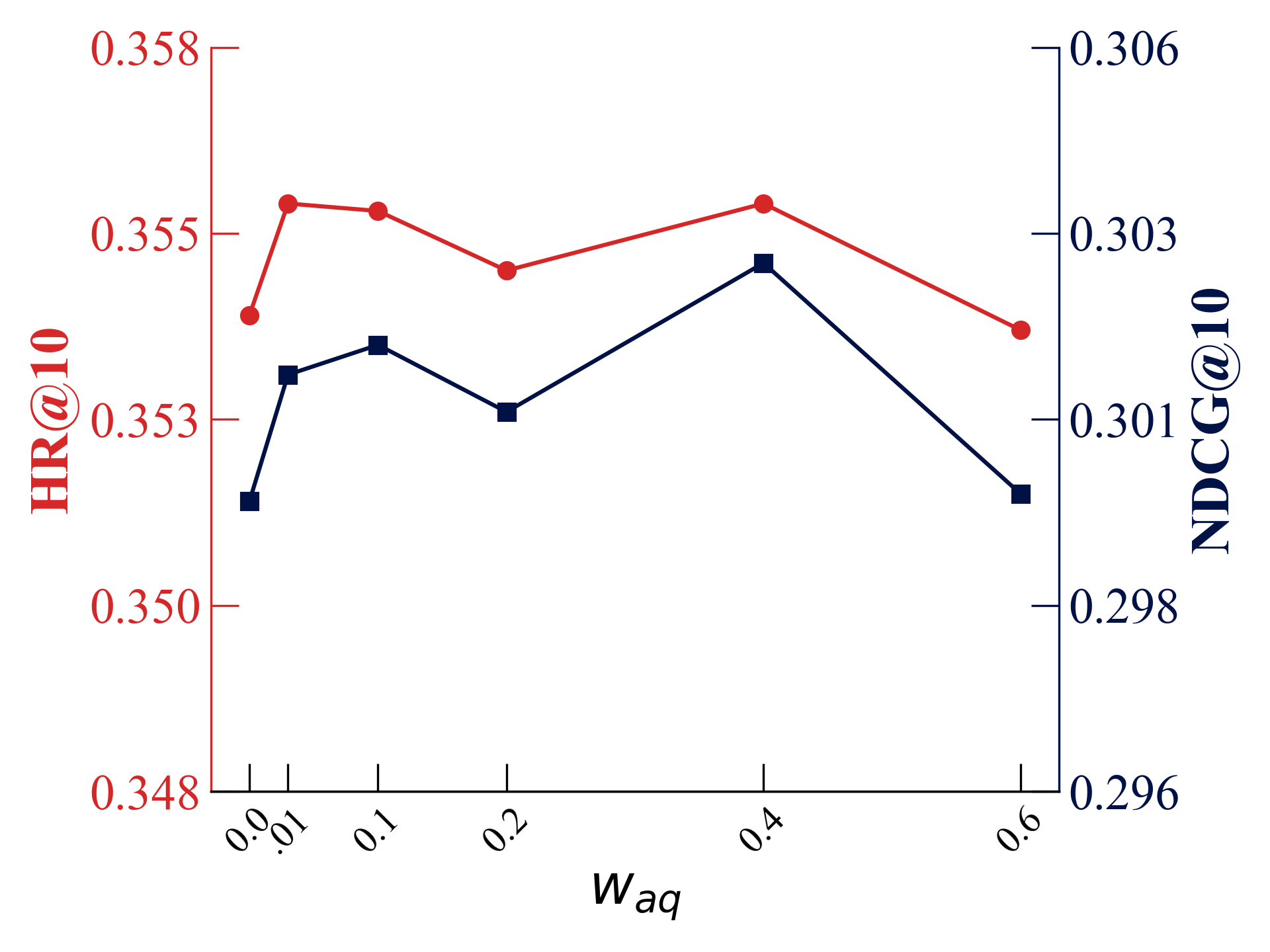}
        \label{effect_waq}
        }
    \end{minipage}
    %\hspace{0.05mm}
    \vspace{-0.75em}
    \caption{ 
    LEASR with different weights of (a) supervised learning and (b) Q-learning augmentation loss on the LFM dataset. }
    % }\vspace{-0.2cm}
    \vspace{-1em}
    \label{effect_wah_waq}
    \vspace*{-4mm}
\end{figure}

Furthermore, we investigate the impact of $w_{ah}$ and $w_{aq}$ 
- the weights of augmentation $\mathcal{L}_{ah}$ and $\mathcal{L}_{aq}$ - in LEASR under the SA2C framework. The results on the LFM dataset are shown in Figures~\ref{effect_wah} and~\ref{effect_waq}, respectively. We can see that the performance initially improves with the increase of $w_{ah}$. The best performance is achieved when $w_{ah}$ is 0.1.
As $w_{ah}$ constantly increases, the performance drops gradually. This suggests that over-weighing augmented actions incrementally diminishes the effect of the true labels, leading to a proportional decrease in performance. We achieve the best performance for $w_{aq}$ = 0.01. As $w_{aq}$ increases, we observe an unstable ascending trend in performance. When beyond a certain threshold, the incremental gains drop. The optimal value of $w_{aq}$ is 0.01. This shows that while the augmentation of Q-learning through predicted action is relatively stable, beyond a certain point further augmentation does not produce proportional performance benefits. 

\subsection{Effect of LE (RQ3)}

\subsubsection{Effect of LE as reward model}
We conduct experiments via the RL-based RS by SA2C framework using the SASRec backbone to see the effect of LE as a reward model (RM). The results on two datasets are reported in Table~\ref{effect_of_rewards}.
LER method replaces the predefined reward of SA2C as the reward return from RM. 
We observe that LER outperforms SA2C across all cases, which substantiates the effectiveness of the RM. We further note that the smaller the $k$ the greater the improvement in the top-$k$ recommendation list, suggesting that a well-designed reward model can effectively prioritize correct actions to higher ranks.
LEA incorporates the augmentation technique into the SA2C, whereas LEAR applies both the augmentation and the reward model. It is evident that LEAR generally outperforms LEA. This superior performance can be attributed to the common advantages gained from the augmentation feedback provided by the LE, which potentially obscures the enhancements derived from the RL components. However, the superiority of LEAR still holds the second phenomenon as described above.

\begin{table*}[h!]
\centering
\caption{Effect of action augmentation on supervised learning (sv) and Q-learning (q). NG is short for NDCG.
Boldface denotes the highest scores.
}
\label{exp_aug_results}
\vspace*{-4mm}
\begin{tabular}{lcp{0.85cm}<{\centering}p{0.85cm}<{\centering}p{0.85cm}<{\centering}p{0.85cm}<{\centering}p{0.85cm}<{\centering}p{0.85cm}<{\centering}cp{0.85cm}<{\centering}p{0.85cm}<{\centering}p{0.85cm}<{\centering}p{0.85cm}<{\centering}p{0.85cm}<{\centering}p{0.85cm}<{\centering}} 
\toprule
\multirow{2}{*}{Model} & &\multicolumn{6}{c}{LFM} & &\multicolumn{6}{c}{Industry}\\
\cmidrule{3-8} \cmidrule{10-15}

 & & HR@5 & NG@5 & HR@10 & NG@10 & HR@20 & NG@20 & & HR@5 & NG@5 & HR@10 & NG@10 & HR@20 & NG@20         \\ 
%\cmidrule{}
\midrule
\multirow{6}{*}{SASRec}  &Normal &0.2798 	&0.2324 &0.3115	&0.2646	&0.3271	&0.2685 &&0.0788 &0.0684&0.1171	&0.0909	&0.1374	&0.0961\\
 &sv &\textbf{0.2899} 	&\textbf{0.2493} &\textbf{0.3263}	&\textbf{0.2757}	&\textbf{0.3408}	&\textbf{0.2793}	&&\textbf{0.0928} &\textbf{0.0759}&\textbf{0.1202}	&\textbf{0.0931}	&\textbf{0.1418}	&\textbf{0.0985}\\
\cmidrule{2-15}
&SA2C  &0.2902 	&0.2501  &0.3372	&0.2851	&0.3516	&0.2885  &   &0.0931 &0.0735&0.1205	&0.0912	&0.1416	&0.0966\\
&sv  &0.3295 &0.2898  &0.3541	&0.2978	&\textbf{0.3755}	&0.3032  &   &0.1026	&0.0847&0.1241    &0.0916	&0.1554 &0.0995\\
&q  &0.3314 &0.2897 &0.3535	&0.2969   &0.3735	&0.3019   &	&0.1022&0.0836&0.1234	&0.0904	&0.1535	&0.0981\\
&sv+q  &\textbf{0.3322}  &\textbf{0.2915}  &\textbf{0.3547}	&\textbf{0.2987}	&0.3754	&\textbf{0.3041} &&\textbf{0.1032} &\textbf{0.0859}&\textbf{0.1252}	&\textbf{0.0929} &\textbf{0.1558}   &\textbf{0.1006}	\\
\bottomrule
\end{tabular}
\end{table*}

\begin{table}[h!]
\centering
\caption{Effect of LE as reward function. NG is short for NDCG.
Boldface denotes the highest score.
}
\label{effect_of_rewards}
\vspace*{-4mm}
\begin{tabular}{lp{0.55cm}<{\centering}p{0.8cm}<{\centering}p{0.8cm}<{\centering}p{0.8cm}<{\centering}p{0.8cm}<{\centering}p{0.8cm}<{\centering}p{0.8cm}<{\centering}} 
\toprule
&SAS  &  HR@5 & NG@5 & HR@10 & NG@10 & HR@20 & NG@20 \\
\midrule
\multirow{4}{*}{{\rotatebox{90}{LFM}}}
& SA2C &0.2902 	&0.2501 &0.3372	&0.2851	&0.3516	&0.2885 \\
& LER &\textbf{0.3252}  &\textbf{0.2714 } &\textbf{0.3401 }	&\textbf{0.2928} &\textbf{0.3539}&\textbf{0.2961}\\
\cmidrule{2-8}
& LEA &0.3296  &0.2802 &0.3405	&0.2877	&0.3548	&0.2912 \\
& LEAR &\textbf{0.3298} &\textbf{0.2833} &\textbf{0.3429}	&\textbf{0.2942}   &\textbf{0.3571}	&\textbf{0.2977}\\
\bottomrule
\multirow{4}{*}{{\rotatebox{90}{Industry}}}
& SA2C &0.0935 &0.0775 &0.1145	&0.0842	&0.1427	&0.0913 \\
& LER &\textbf{0.0994} & \textbf{0.0841} &\textbf{0.1219}&\textbf{0.0928}	&\textbf{0.1432}	&\textbf{0.0982}	\\
\cmidrule{2-8}
& LEA &0.1032 &0.0859&\textbf{0.1252}&\textbf{0.0929} &\textbf{0.1558}  &\textbf{0.1006} \\
& LEAR &\textbf{0.1037}&\textbf{0.0861}&0.1231&0.0922&0.1525&0.0996\\
\bottomrule
\end{tabular}
\end{table}

\begin{table}[h!]
\setlength{\tabcolsep}{4.8pt}
\centering
\caption{Effect of LE as state model. NG is short for NDCG.
Boldface denotes the highest score.
}
\label{effect_of_state}
\vspace*{-4mm}
\begin{tabular}{p{0.35cm}p{0.55cm}<{\centering}p{0.8cm}<{\centering}p{0.8cm}<{\centering}p{0.8cm}<{\centering}p{0.8cm}<{\centering}p{0.8cm}<{\centering}p{0.8cm}<{\centering}} 
\toprule
\multicolumn{2}{c}{\multirow{2}{*}{ Model}} & \multicolumn{6}{c}{LFM} \\
\cmidrule{3-8}
 & & HR@5 & NG@5 & HR@10 & NG@10 & HR@20 & NG@20 \\
\midrule
\multirow{3}{*}{\shortstack{SAS} } & SA2C&0.2902 &0.2501 &0.3372	&0.2851	&0.3516	&0.2885 \\
& LES$'$ &0.3202 &0.2804 & 0.3504 & 0.2956 & 0.3711 &0.3008\\
& LES &\textbf{0.3311} &\textbf{0.2913}& \textbf{0.3539} & \textbf{0.2988} &\textbf{0.3734} & \textbf{0.3037}\\
\bottomrule
\multirow{3}{*}{\shortstack{GRU}}  & SA2C &0.2896&0.2743&0.3173&0.2797&0.3337&0.2838  \\
& LES$'$ &0.2973 &0.2597 & 0.3141 & 0.2641 & 0.3345 &0.2785\\
& LES &\textbf{0.3128} &\textbf{0.2818} &\textbf{0.3321} &\textbf{0.2891} & \textbf{0.3518} & \textbf{0.2935} \\
\bottomrule
\end{tabular}
\end{table}

\subsubsection{Effect of LE as state model}
We conduct experiments within the SA2C framework on the LFM dataset between three methods: (1) the baseline SA2C, relying on the hidden state from the sequential model, i.e., $h_t$ in Eq.~\ref{le_s}, (2) LES$'$ that uses the state from SM, i.e., ${s'}^e_t$ in Eq.~\ref{le_s'}, and (3) LES that leverages both the state from LE and hidden state from the sequential model i.e., ${s}^e_t$ in Eq.~\ref{le_s}. Table~\ref{effect_of_state} indicates that LES has the highest performance across all backbones and metrics, while LES$'$ outperforms SA2C in most cases. This demonstrates that leveraging effectively all state representations can introduce performance gains in RL-based RS.

\begin{figure}[h!]
    \captionsetup[subfloat]%{captionskip=-5pt,nearskip=0pt,farskip=0pt}
    {}
    \centering
    \begin{minipage}[t]{0.23\textwidth}
     \centering
        \subfloat[]{%
        \includegraphics[trim={0 0 0 0}, clip, scale=0.22]{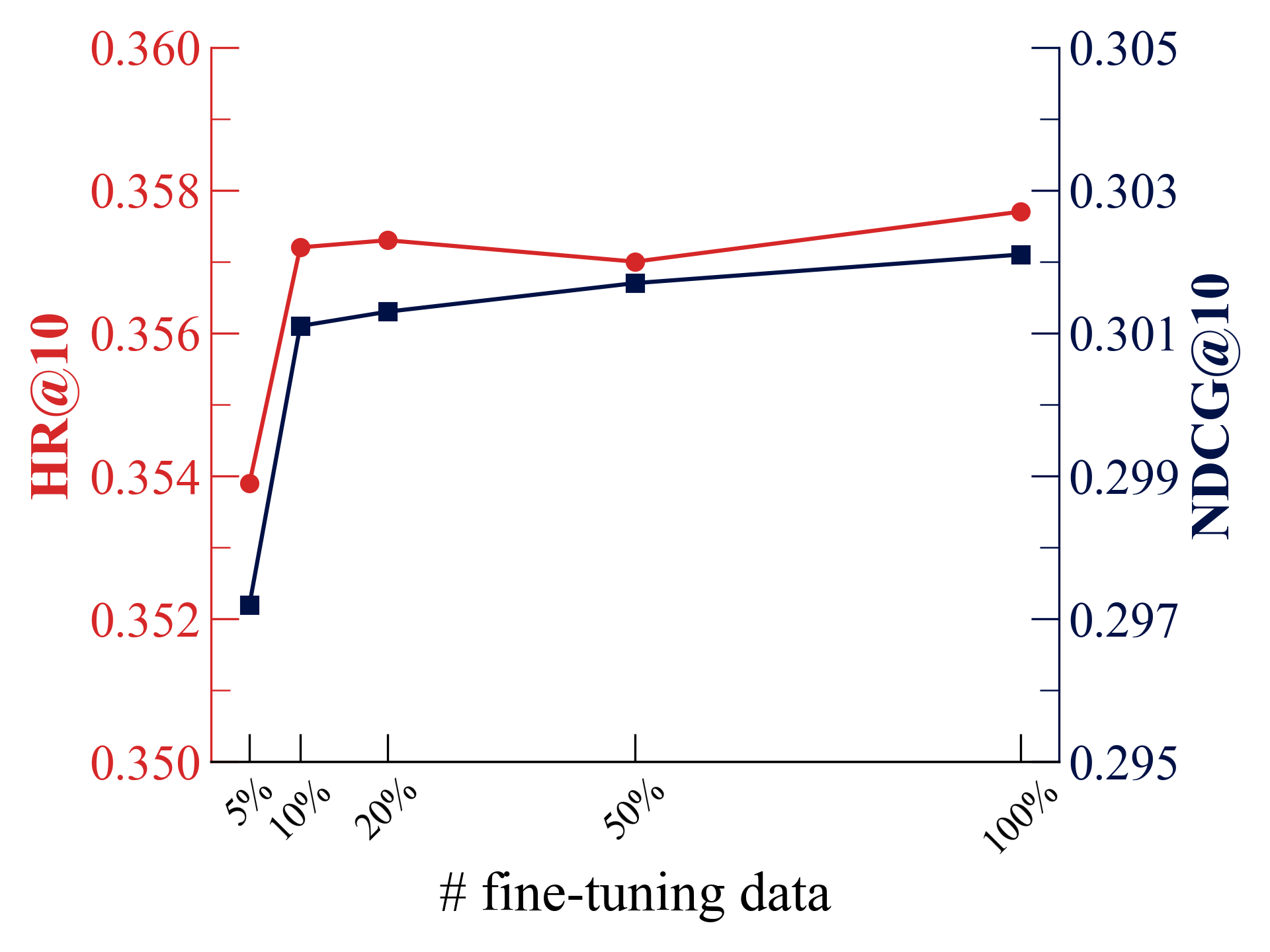}
        \label{scaling_leasr}
        }
    \end{minipage}
    \hspace{0.07mm}
    %\hfill
    \begin{minipage}[t]{0.23\textwidth}
     \centering
        \subfloat[]{%
        \includegraphics[trim={0 0 0 10}, clip, scale=0.22]{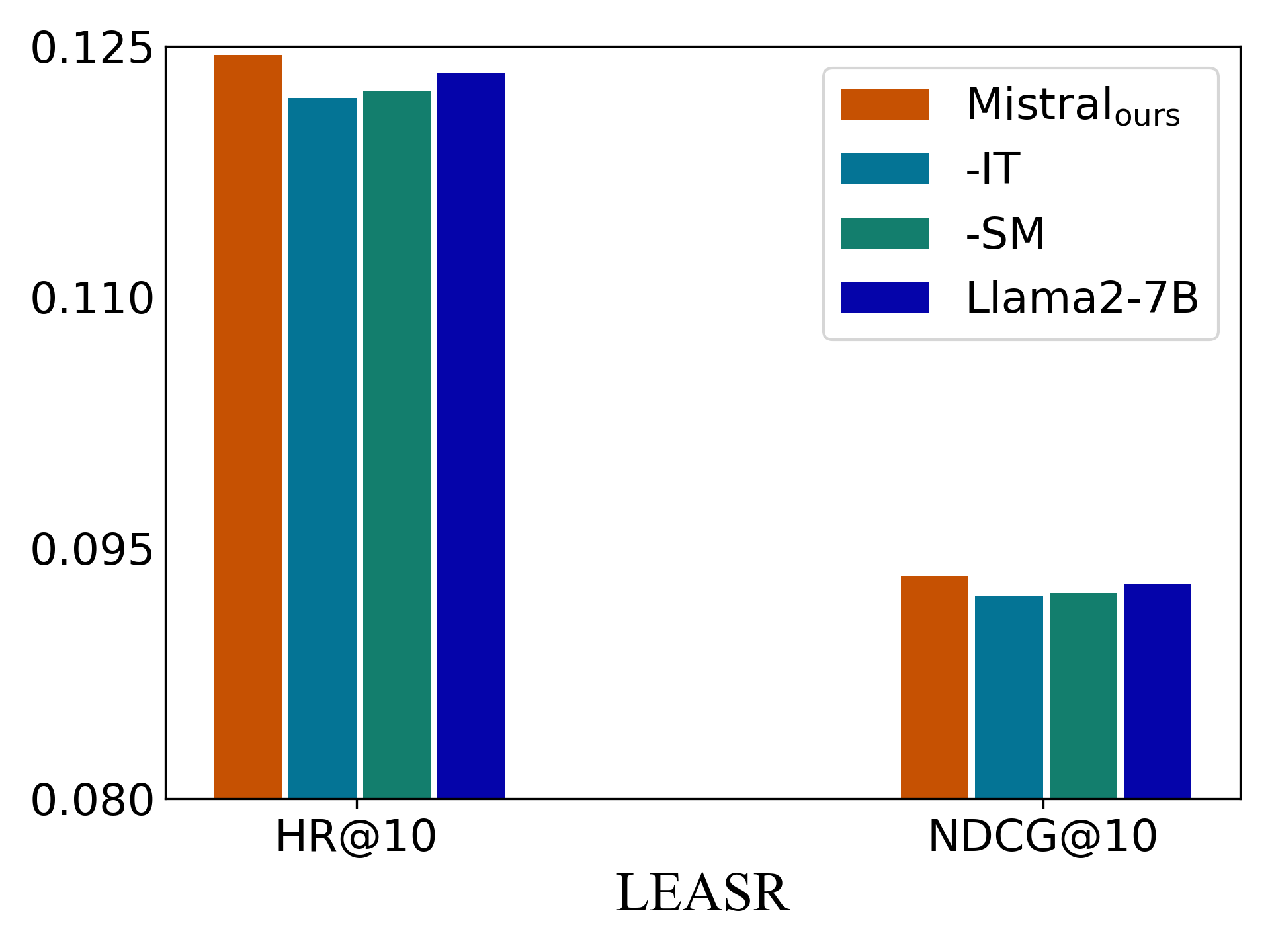}
        \label{effect_le_it_sm_llm}
        }
    \end{minipage}
    %\hspace{0.05mm}
    \vspace{-0.75em}
    \caption{ 
    Effect of scaling the training data for LE. The result of LEASR on the LFM dataset. }
    % }\vspace{-0.2cm}
    \vspace{-1em}
    \label{effect_le}
\end{figure}

\subsection{Effect of learning strategies for LE (RQ4)} 

\subsubsection{Effect of data scale.}
Despite optimizing the LE on a subset of the training data, we study the effect of data scaling to provide insights into how to more efficiently obtain LE in the future. We train the LE on a portion of the LFM dataset with varying portions: 5\%, 10\%, 20\%, 50\%, and 100\%. The results on the LEASR method are shown in Figure~\ref{scaling_leasr}.
These indicate that as the data portion increases to 10\%, we observe a notable enhancement in performance, demonstrating that a bigger dataset enables the LE to capture complex user-item interactions and return effective user feedback. The scale we set is 10\%. As the scale continuously increases, the effect of LE grows slowly and fluctuates within a range. This demonstrates that our method is capable of efficiently optimizing LLMs with a small data portion, producing a state, reward model, and augmented feedback that are scalable to the entire user data. 

\subsubsection{Effect of item tokenization
}

We compare the performance of LE utilizing item tokenization against representing items by their textual content (-IT), where user-item interactions are generated from sequences of item titles. We utilize LEASR to examine the impact on LE. Figure~\ref{effect_le_it_sm_llm} shows the results, which suggest that applying item tokenization surpasses the concatenation method. 
This underscores that the items tokens,  represent the semantic information in a concise way, and hence contribute to an effective learning of interaction sequences. 
\subsubsection{Effect of state loss
}
To observe the impact of the loss $\mathcal{L}^e_{sm}$ on state generation,
we train LE without (-SM) contrastive learning and show the impact by the LES method.
Figure~\ref{effect_le_it_sm_llm} shows a decrease in performance when 
$\mathcal{L}^e_{sm}$ is omitted. This implies that the SM loss plays a critical role in refining the state representation, which is integral to the robustness and accuracy of the LE. 

\subsubsection{Effect of LLMs}
We show the effect of different LLMs on LE by comparing Mistral with Llama2-7B, a model with a 7B parameter count similar to Mistral but launched earlier. Figure~\ref{effect_le_it_sm_llm} shows that the performance of Mistral is slightly higher than Llama2-7B. We argue that as LLMs evolve, there is the potential for continuous improvement in building LE.

\section{Conclusion and Future Work}

In contrast to adaptations of LLMs used as recommenders, the inclusion of LE enhances RL-based sequential models without requiring significant increases in inference computations, hence our model is very easy to deploy in real world recommendation settings. 
The use of LLM's as reward, state and action models is shown to be a promising direction in the context of RL-driven recommender systems. Future directions of research can include the shaping of different rewards and the inclusion of additional user behavior data in the LLM to create better state representations. We also intend to investigate alternative fine-tuning methods for both reward state and action generation. Moreover, the use of more powerful LLM's can lead to even better results.  

%\eject 
\bibliographystyle{ACM-Reference-Format}
\bibliography{sample-base}

\end{document}